\documentclass{ws-procs9x6}
\usepackage{amsmath}
\usepackage{graphicx, psfrag}
\usepackage{amsmath}
\usepackage{amsfonts}
\usepackage{amssymb}  
\usepackage{fancybox}
\DeclareMathOperator{\tgh}{th}
\DeclareMathOperator{\ch}{ch}
\DeclareMathOperator{\sh}{sh}
\newcommand{\pn}{\par\noindent}

\begin{document}

\title{QUANTUM SPIN CHAINS AT FINITE TEMPERATURES}
\author{Frank G{\"o}hmann}
\address{Fachbereich C -- Physik, Bergische Universit\"at Wuppertal,\\
42097 Wuppertal, Germany}

\author{Junji Suzuki}
\address{Department of Physics, Faculty of Science, Shizuoka University,\\
Ohya 836, Suruga, Shizuoka, Japan}

\begin{abstract}
This is a pedagogical review on recent progress in the exact
evaluation of physical quantities in interacting quantum systems
at finite temperatures. 1D quantum spin chains are discussed in detail as
 typical examples.

\end{abstract}

\keywords{Quantum Transfer Matrix; correlation functions at finite
temperatures.}

\bodymatter

\section{Introduction}

The
evaluation of the thermal average of physical quantities
is one of the main aims in statistical mechanics. 
The density matrix of a system is the most fundamental quantity 
to achieve this aim.
Its diagonalization, however, becomes exponentially difficult
with growing system size $L$. One inevitably has to give up this procedure
in the thermodynamic limit.
An alternative approach for quantum systems 
is to diagonalize the Hamiltonian, and to sum up the
contributions from each eigenstate.
This means to divide the problem into two parts: (1) diagonalize,
and (2) sum up. Again,
both procedures become exponentially difficult with the increase
of $L$.

In this article we re-consider this problem for integrable quantum spin
chains. We will show how the integrability helps bypassing the difficulties
and yields exact estimates.
The first problem, the diagonalization of the Hamiltonian, can, in
principle, be solved by the celebrated Bethe ansatz. The second step,
however, remains as a cliff wall.
A first breakthrough, the string hypothesis approach,
was achieved in the early 70's \cite{Gaudin71,TakSuz}.
%
%
In this approach
one introduces so-called root density functions of strings and holes of
various lengths for the diagonalization. The free energy becomes
a functional of these density functions,
which is claimed to be
%
exact near
its minimum. Therefore the variational estimate (w.r.t. density functions), 
with a fixed energy of the system, 
yields the exact free energy.
The string hypothesis formulation can be regarded as a micro-canonical
approach. 
It is supported by many
consistency tests. We conclude that within the string hypothesis approach
the diagonalization is achieved, but the summation is cleverly avoided.

In order to evaluate thermal expectation values of operators, it is better
to deal with the canonical ensemble.
We therefore consider an alternative approach based on the Quantum Transfer
Matrix (QTM)\cite{Suz85, SuNaWa92}. It utilizes an exact mapping between
a 1D quantum system at finite temperatures and a 2D classical system.
At first sight
the formulation may look tautological 
and may seem to be suffering from the need of ``summation".
Yet,
the main claim of the QTM formulation is that this is not the case. As in
the the string hypothesis approach the ``summation'' can be avoided.
Moreover, the QTM  makes the evaluation of many quantities of physical
relevance straightforward. 

This article is organized as follows. In Sec.\ 2, we present a review on
the QTM formulation. The results for the bulk quantities will be
summarized in Sec.\ 2.3.
In the rest of Sec.\ 2, we supplement arguments to justify the
formula in Sec.\ 2.3. The non-linear integral equation (NLIE)  will be explained in Sec.\ 3 together with
an example for the explicit evaluation of bulk quantities.
The evaluation of the reduced  density matrix elements (DME)  will be discussed in Sec.\ 4.
\section{The QTM formulation}
\subsection{The problem}
Let ${\cal H}$ be the Hamiltonian of a 1D quantum system of size $L$ and
$V$ its space of states. Our goal is to calculate the thermal expectation
value of any physical quantity ${\cal O}$ at temperature $T (=1/\beta)%
$\footnote{The Boltzmann constant $k_B$ is set to be unity in this report.}
in the limit $L \rightarrow \infty$,
\begin{equation}
\langle   {\cal O} \rangle =  \lim_{L \rightarrow \infty} \frac{ {\rm tr}_V \, {\cal O} \,{\rm e}^{-\beta {\cal H}}}{Z_{\rm 1D}(\beta) }  
\qquad 
Z_{\rm 1D}(\beta)  ={\rm tr}_V \, {\rm e}^{-\beta {\cal H}} =\sum_{j}  {\rm e}^{-\beta  E_j}.   \label{defZ}
\end{equation}
Here $E_j$ stands for an eigenvalue of  ${\cal H}$.

The definition requires both  diagonalization and  summation.
Below we shall show how we can avoid the latter within the framework
of QTM.


\subsection{The Baxter-L{\"u}scher formula}
To be concrete, we specify a Hamiltonian. As a prototypical integrable
lattice system we choose the 1D spin $\frac{1}{2}$ XXZ model,
\begin{equation}
\label{hamxxz}
{\cal H}= J \sum_{j=1}^L  
\Bigl(\sigma^x_j \sigma^x_{j+1} + \sigma^y_j \sigma^y_{j+1} + \Delta (\sigma^z_j \sigma^z_{j+1}+1) \Bigl) 
  =\sum_{j=1}^L  \hat{h}_{j, j+1}
\end{equation}
where the $\sigma^a\, (a=x,y,z)$ are the Pauli matrices. The  periodic boundary conditions (PBCs) imply
$\sigma^a_{L+1}=\sigma^a_1$. The anisotropy is parameterized as $\Delta
= \cos \gamma$.  
The Hamiltonian acts on ``the physical space" $V_{\rm phys}:=
\bigotimes_{j=1}^L V_j$ where $V_j$ denotes the $j$th copy of a two-%
dimensional vector space $c_1 {\bf e}_+ +c_2  {\bf e}_- $.
The trace in (\ref{defZ}) must be performed over $V_{\rm phys}$.
By definition the ``Hamiltonian density" $\hat{h}_{j,j+1}$ is the $j$th
summand in the first sum in (\ref{hamxxz}). It acts non-trivially only
on $V_{j} \otimes V_{j+1}$.
 
\begin{figure}
\psfrag{u}{$u$}
\psfrag{v}{$v$}
\psfrag{[u-v]}{$[\frac{u-v}{2}]$}
\psfrag{[u-v+1]}{$[\frac{u-v}{2}+1]$}
\psfrag{quv}{$q^{\frac{u-v}{2}}$}
\psfrag{a}{$\alpha$}
\psfrag{b}{$\beta$}
\psfrag{g}{$\gamma$}
\psfrag{d}{$\delta$}
\psfrag{Rabgd}{$R_{\beta  \delta} ^{\alpha \gamma}(u,v)$}

\begin{center}
     { \includegraphics[width=8cm]{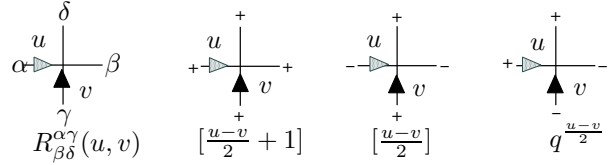}}
\end{center}
\caption{A graphic representation for $R_{\beta  \delta} ^{\alpha \gamma}(u,v)$ and some examples }
\label{sixv_fig}
 \end{figure}

The above Hamiltonian is integrable in the following sense. Let $R(u,v)$
be the $U_{q}(\widehat{\mathfrak{sl}_2})$ $R$ matrix\cite{Jimbo},
$$
R(u,v) = 
\begin{pmatrix}
[1+\frac{u-v}{2}]&                &               &                & \\
&   [\frac{u-v}{2}]&            q^{\frac{-u+v}{2}}&       &\\
&     q^{\frac{u-v}{2}}&        [\frac{u-v}{2}]&       &\\
  &               &                 &                    [1+\frac{u-v}{2}]&
\end{pmatrix}
\quad [u] := \frac{q^{u} -q^{-u}}{q-q^{-1}}
$$
depending on the spectral parameters (or rapidities) $u, v\in \mathbb{C}$. 
We define  $E^{\alpha}_{\beta}$ s.t.
$(E^{\alpha}_{\beta})_{i,j}= \delta_{\alpha,i} \delta_{\beta,j}$.
Then the matrix elements $ R_{\beta  \delta} ^{\alpha \gamma} $ can be read
off from
$$
R(u,v) = \sum_{\alpha,\beta, \gamma, \delta=1,2}    R_{\beta  \delta} ^{\alpha \gamma}(u,v)  
E^{\beta}_{\alpha} \otimes E^{\delta}_{\gamma}.
$$
The index $1 (2)$ refers to ${\bf e}_+  ({\bf e}_-)$. See
fig.~\ref{sixv_fig} for a graphic representation.
We put arrows, to distinguish the $R$ matrix from other $R$ matrices
appearing below. The reader should not confuse them with physical
variables.

By  $R_{j,j+1}(u,v)$ we mean the $R$ matrix acting non-trivially only
on the tensor product $V_{j}(u) \otimes V_{j+1}(v)$ of
$U_{q}(\widehat{\mathfrak{sl}_2})$ modules. We also introduce the intertwiner
$R^{\vee}_{j,j+1}(u, v) = P_{j,j+1} R_{j,j+1}(u,v)$, where 
$P:  V_{j}(u) \otimes V_{j+1}(v) \rightarrow   V_{j+1}(v) \otimes V_{j}(u)$.
Then, with $q={\rm e}^{i \gamma}$, we have the expansion
$$
 R^{\vee} _{j,j+1}(u,0) = 1+  \frac{\gamma}{4J \sin \gamma} u \, (\hat{h}_{j,j+1} 
 + \hat{h'}_{j,j+1}   ) 
 +O(u^2),
$$
where  $\hat{h'}_{j,j+1} :=iJ \sin \gamma (\sigma^z_j- \sigma^z_{j+1})$.
We introduce the row-to-row (RTR) transfer matrix $T_{\rm RTR}(u)
\in {\rm End}( V_{\rm phys})$,
\begin{equation}
T_{\rm RTR}(u) = {\rm tr}_{a} R_{a,L}(u,0)   R_{a,L-1}(u,0)  \cdots   R_{a,1}(u,0).
\label{defRTR}
\end{equation}
With the lattice translation operator ${\rm e}^{i P}$, shifting the state
by one site, we obtain the Baxter-L{\"u}scher formula\cite{Bax8v}
\begin{equation}
T_{\rm RTR}(u)  = {\rm e}^{i P} \bigl(1+ \frac{\gamma u}{4J \sin \gamma} \, {\cal H} + O(u^2) \bigr).
\label{RTRandH}
\end{equation}
Note that the $\hat{h'}_{j,j+1}$ terms cancel due to the PBCs. The huge
symmetry $U_{q}(\widehat{\mathfrak{sl}_2})$ is at the bottom of
the integrability of the Hamiltonian.


\subsection{A summary of results for bulk quantities}
We first present the formula for the free energy per site
in the QTM formalism. 
A supplemental discussion will be given in subsequent sections.

\begin{figure}
\psfrag{u}{$u$}
\psfrag{v}{$v$}
\psfrag{[v-u]}{$[\frac{v-u}{2}]$}
\psfrag{[v-u+1]}{$[\frac{v-u}{2}+1]$}
\psfrag{qvu}{$q^{\frac{u-v}{2}}$}
\psfrag{a}{$\alpha$}
\psfrag{b}{$\beta$}
\psfrag{g}{$\gamma$}
\psfrag{d}{$\delta$}
\psfrag{tRabgd}{\small{$(R^{t})^{\alpha \gamma}_{\beta \delta}(u,v)$}}
\begin{center}
     { \includegraphics[width=7cm]{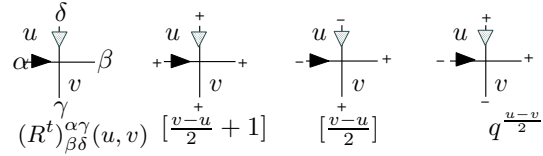}}
\end{center}
\caption{A graphic representation for $(R^{t})^{\alpha \gamma}_{\beta \delta}(u,v)$ and some examples }
\label{sixvt_fig}
 \end{figure}

We introduce the transposed $R$ matrix $R^{t}_{j,k}(u,v)$\cite{Klu93} by
$
(R^{t})^{\alpha \gamma}_{\beta \delta}(u,v) = R^{\delta \alpha}_{\gamma \beta}(v,u).
$
See fig.~\ref{sixvt_fig}.
The QTM  does not act on $V_{\rm phys}$ but on a fictitious space 
$V_{\rm Trotter}=V_1(u) \otimes V_2(-u) \otimes \cdots V_{N-1} (u)
\otimes  V_{N} (-u)$. The fictitious system size $N$ is often referred
to as the Trotter number. The parameter $u$ is fixed to be 
$$
u=-\frac{ 4\beta J \sin \gamma }{ \gamma N} =-\frac{4J \sin \gamma  }{\gamma N T}.
$$

In its most sophisticated version, the QTM is explicitly defined by
\cite{Klu93},
\begin{equation}
\mspace{-0.5mu}
T_{QTM} (x, u) ={\rm tr}_a  R_{a N}(ix, -u) R^{t}_{a,N-1} (ix,u) \cdots 
R_{a2}(ix, -u) 
R^t_{a1}(ix, u).
\label{defQTM}
\end{equation}
The new parameter $x$ will later play the role of a spectral parameter. 
The factor $i$ is introduced for convenience.

We are now in a position to write down the formula for the free energy
per site in the  thermodynamic limit, $f = - \lim_{L \rightarrow \infty}
\frac{T}{L} \ln Z_{{\rm 1D}}(\beta)$.
\begin{theorem}\label{Suzukitheorem}
Let $\Lambda_0 $ be the largest eigenvalue of $T_{QTM}(0, u)$.
Then the free energy per site is solely given by $\Lambda_0 $,
\begin{equation}
f= - \lim_{N \rightarrow \infty}  T \ln  \Lambda_0.
\label{formulaf}
\end{equation}
\end{theorem}

The limit  $N \rightarrow \infty$ is referred to as the Trotter limit.
As was announced earlier, eq.\ (\ref{formulaf}) expresses $f$ {\it without
recourse to any summation}. We also note that  $\ln \Lambda_0$ itself
is already intensive, which may reflect the size dependent interaction
of the system.

The quantitative analysis of (\ref{formulaf}) is most efficiently performed
by means of the NLIE.  
Having in mind the
examples,  from now on we are considering 
 only for
$J=\frac{1}{4}, \gamma \rightarrow 0$ and $u=-\frac{\beta}{N}$, consequently.
Let $\mathfrak{a}$ be the unique solution to the NLIE\footnote{To be precise, there are, in general, several
equivalent versions of NLIEs. We present one of these below.}
\begin{align}
\ln \mathfrak{a}(x)  &= \beta \epsilon_0(x+i) -\int_{\cal C} \frac{2}{ (x-y)^2+4}  \ln \mathfrak{A}(y)\frac{dy}{\pi}
\label{NLIE}  \\
\epsilon_0(x)&=h+\frac{2}{(x-i)(x+i)} \qquad 
\mathfrak{A}:=1+\mathfrak{a}.   \nonumber 
\end{align}
Here the contour ${\cal C}$ is a closed narrow loop which encircles all
``Bethe roots". 
We added a Zeeman term $\frac{h}{2} \sum_j \sigma^z_j$ to the
Hamiltonian so that ${\rm diag}(\exp(-\frac{\beta h }{2}), 
\exp(\frac{\beta h}{2}))$ is inserted in the trace in (\ref{defQTM}).
Then we have the following
\begin{theorem}\label{theoremf}
The free energy per site can be evaluated in terms of the solution to the
NLIE.
\begin{equation}
\beta f= \frac{\beta}{2}( 1+h ) 
-\int_{\cal C}  \frac{1}{x(x+2 i)} \ln \mathfrak{A}(x) \frac{dx}{\pi}. 
\label{freeA}
\end{equation}
\end{theorem}

Note that the NLIE (\ref{NLIE}) and the expression for $f$ in (\ref{freeA})
are independent of $N$. The extension to arbitrary $J, \gamma$ is
straightforward.

Below we shall comment on the derivation of the formula.
By presenting supplementary arguments, we wish to convince the reader 
that the above formalism, seemingly complicated, is actually necessary
and efficient for many purposes.
Hereafter we set again $h=0$ for simplicity.
\subsection{The 1D quantum partition function as a 2D classical 
partition function}

\begin{figure}
\psfrag{u}{$u$}
\psfrag{v}{$v$}
\psfrag{[v-u]}{$\frac{v-u}{2}$}                         
\psfrag{[v-u+1]} {$\frac{v-u}{2}+1$}                   
\psfrag{qvu}{$1$}                             
\psfrag{a}{$\alpha$}
\psfrag{b}{$\beta$}
\psfrag{g}{$\gamma$}
\psfrag{d}{$\delta$}
\psfrag{tRabgd}{$\widetilde{R}^{\alpha \gamma}_{\beta \delta}(u,v)$}

\begin{center}
     { \includegraphics[width=8cm]{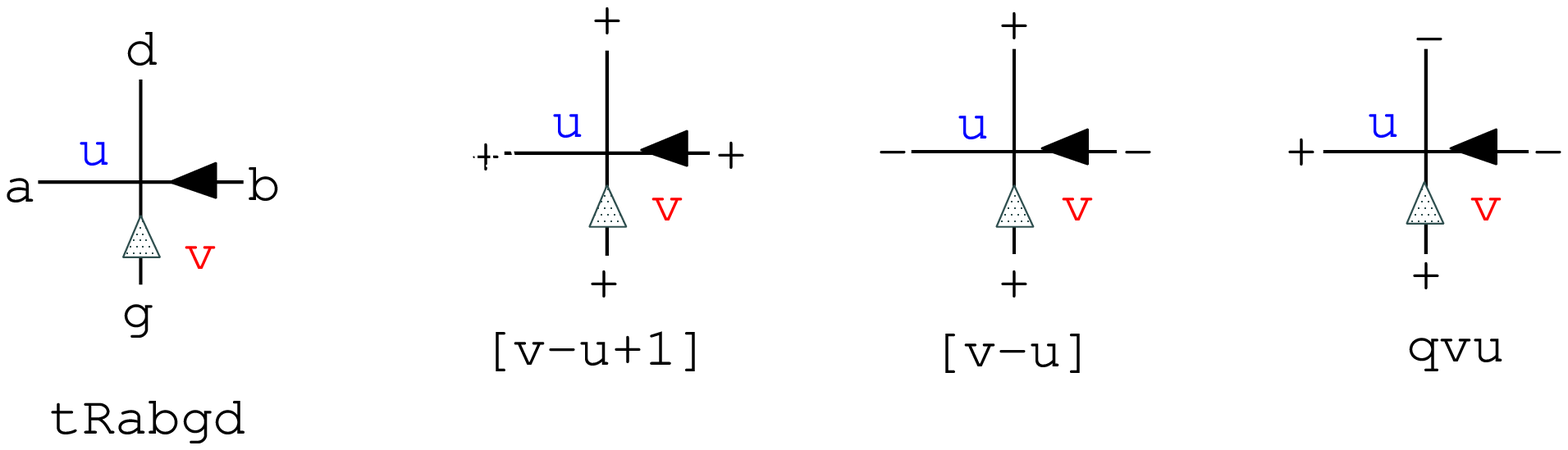}}
\end{center}
\caption{A graphic representation for
$\widetilde{R}^{\alpha \gamma}_{\beta \delta}(u,v)$ and some examples}
\label{sixvr_fig}
\end{figure}

\begin{figure}
\psfrag{u}{$u$}
\psfrag{-u}{$-u$}
\psfrag{v}{$v$}
\psfrag{N}{$N$}
\psfrag{L}{$L$}
\psfrag{0}{$0$}
\begin{center}
     { \includegraphics[width=4cm]{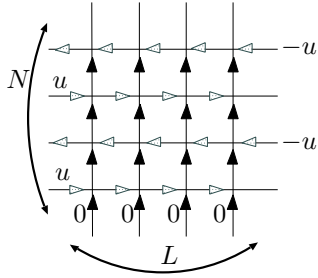}}
\end{center}
\caption{Fictitious two-dimensional system}
\label{twoDz}
 \end{figure}

We define a rotated R matrix $\widetilde{R}(u,v)$ by 
$ \widetilde{R}^{\alpha \gamma}_{\beta \delta}(u,v) =
R^{\gamma \beta}_{\delta \alpha} (v,u)
$ (fig.~\ref{sixvr_fig}).
Then we introduce a rotated transfer matrix $\widetilde{T}_{\rm RTR}(u)
\in {\rm End}( V_{\rm phys})$ by
$$
\widetilde{T}_{\rm RTR}(u) = {\rm tr}_{a}\widetilde{R}_{a,L}(-u,0)  \widetilde{R}_{a,L-1}(-u,0)  \cdots  \widetilde{R}_{a,1}(-u,0).
$$
Analogous to (\ref{RTRandH})
$
\widetilde{T}_{\rm RTR}(u)  = {\rm e}^{-i P} \bigl(1+ u \, {\cal H} + O(u^2) \bigr).
$
%
We thus obtain an important identity,
\begin{equation}
Z_{\rm 1D}(\beta) = {\rm tr}_{V_{\rm phys}} {\rm e}^{-\beta {\cal H}}
=\lim_{N \rightarrow \infty} 
{\rm tr}_{V_{\rm phys}}  \bigl(T_{\rm double}(u) )^{\frac{N}{2}}  |_{u \rightarrow -\frac{\beta}{N}}
\label{equivalence1}
\end{equation}
where $T_{\rm double}(u):= T_{\rm RTR}(u) \widetilde{T}_{\rm RTR}(u) $.
The rhs of (\ref{equivalence1}) can be interpreted as a partition function
of a 2D classical system defined on $N \times L$ sites
(fig.~\ref{twoDz}), 
$$
Z_{\rm 1D}(\beta) = \lim_{N \rightarrow \infty}   Z_{\rm 2D classical} (N, L, u=-\frac{\beta}{N} ).
$$

\begin{figure}[hbtp]
\centering
{  \includegraphics[width=4cm]{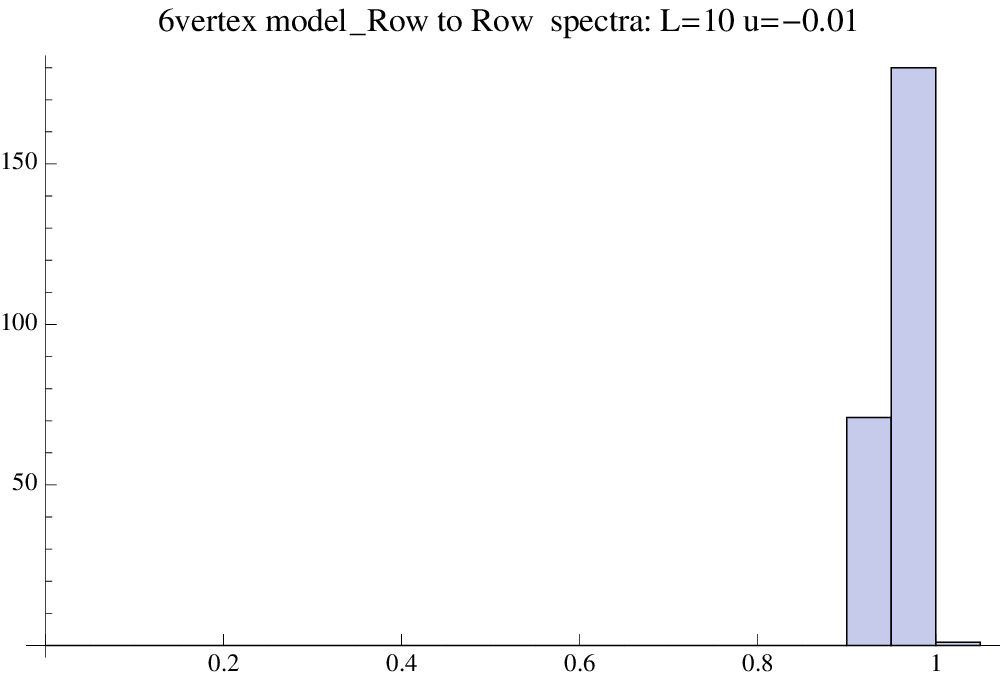} \hspace{1cm}
  \includegraphics[width=4cm]{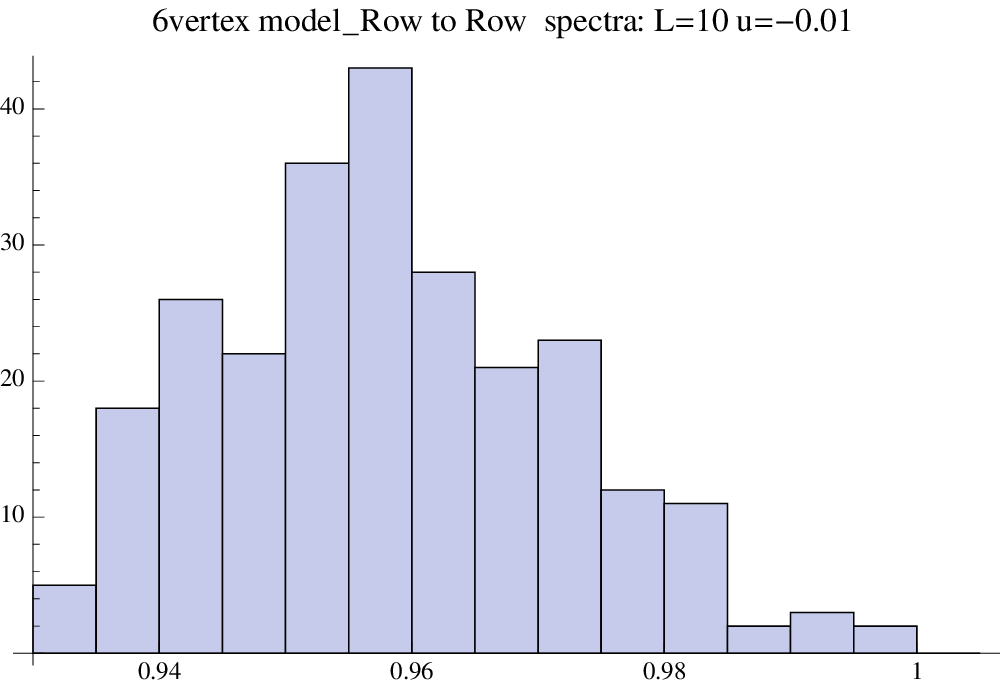} }
\caption{The distribution of eigenvalues. The horizontal axis is the
absolute value of the eigenvalues normalized by the largest one. The left
figures ranges over [0,1] in the horizontal direction, and the right one is
zoomed into the range [0.93,1].
}
\label{afig}
\end{figure}

This equivalence lies in the heart of the QTM formalism. The expression
(\ref{equivalence1}) itself, however, is of no direct use for the actual
evaluation of physical quantities for the following reason.
Let the eigenvalue spectrum of  $T_{\rm RTR}(u)$ be $\lambda_0(x) >
\lambda_1(x) \ge \lambda_2(x)  \ge \cdots $. We introduced $x=i^{-1}(u+1)$
for technical reasons. It is easy to see that $\widetilde{T}_{\rm RTR}(u)$
has the same spectrum. Thus,
\begin{equation}
{\rm tr}_{V_{\rm phys}}  \bigl(T_{\rm double}(u)  \bigr)^{\frac{N}{2}}=
\bigl(\lambda_0(x) \bigr)^{N} \Bigl( 1+  \bigl( \frac{\lambda_1 (x) }{\lambda_0 (x) }\bigr )^{N}  + 
 \bigl( \frac{\lambda_2(x) }{\lambda_0(x) }\bigr )^{N}
  +\cdots \Bigr ).
  \label{RTRexpansion} 
\end{equation}
The eigenvalue $\lambda_j (x)$ is characterized by its zeros
$\pm \theta_a\, (a=1,2,\dots) $ on the real axis
(holes).
We know numerically  that for low excitations, $\theta_a \sim \ln L$
and also that $\lambda_j (x)$ is analytic and nonzero in the strip 
$|\Im {\rm m} x | \le 1$ except at $\pm \theta_a$.
Let us introduce an analytic and nonzero function near the real axis,
${\lambda}^{\sharp}_j $, by $\lambda_j(x)= \prod_a \tgh \frac{\pi }{4}
(x-\theta_a) \tgh \frac{\pi }{4}(x+\theta_a)  {\lambda}^{\sharp}_j (x) $. 
It approximately satisfies the inversion relation for  $L \gg 1$, 
\begin{equation}
{\lambda}^{\sharp}_j (x-i) {\lambda}^{\sharp}_j (x+i) = \phi(x),
\label{inversion}
\end{equation}
where $\phi(x)$ is a known function common to any $j$. Thus, we simply
have
$$
\left |\frac{\lambda_j(x)}{ \lambda_0(x)   } \right |  =\left|\prod_a \tgh \frac{\pi }{4} (x-\theta_a) \tgh\frac{\pi }{4}(x+\theta_a) \right|.
$$
For very low excitations, we take a single pair of holes, substitute
$\theta_a \sim \frac{2}{\pi} \ln \frac{2 \pi L}{\Delta_j}$ and take the
large $L$ limit. Then we arrive at the estimate ($u \sim 0$)
\begin{equation}
\left| \frac{\lambda_j(x)}{ \lambda_0(x)   } \right |  \sim    {\rm e}^{ - \frac{|u|\Delta_j}{L}} \quad{\rm thus} \quad 
\left| \frac{\lambda_j(x)}{ \lambda_0(x)   }  \right |^{N}  \sim    {\rm e}^{ - \Delta_j \frac{N}{L}|u| }.
\label{estimateRTR}
\end{equation}
For a usual 2D classical system we can consider an infinitely long
cylinder and take $\frac{N}{L} \gg 1$. We thus have to take into account
only the first term on the rhs in (\ref{RTRexpansion}).
By contrast, the spectral parameter depends on the fictitious system size 
$u=-\frac{\beta}{N}$ in the present case. Therefore, as long as $T \ne 0$,
we have 
$$
\left |\frac{\lambda_j(x)}{ \lambda_0(x)   }\right |^{N}  \sim    {\rm e}^{ - \Delta_j \frac{\beta}{L}}  = O(1)  \quad {\rm for}\,\,  L \gg 1.
$$
Fig.~\ref{afig} presents numerical evidence for the above argument.
The left figure shows the histogram of the distribution of
$|\lambda_j/\lambda_0|$ for $q=1,  L=10, u=-0.01$ in the sector with vanishing magnetization.
One clearly sees that the maximum of the distribution lies near
$|\lambda_j/\lambda_0 | \sim 1$. The right figure magnifies the region
near $|\lambda_j/\lambda_0| \sim 1$. The maximum is located around
$|\lambda_j/\lambda_0| \sim 0.96$. We believe that, with increasing $L$,
the peak moves towards  $|\lambda_j/\lambda_0| \sim 1$. These findings
are consistent with (\ref {estimateRTR}).
Hence, we conclude that infinitely many terms of the sum in the rhs of
(\ref{RTRexpansion}) contribute non-trivially, and eq.\
(\ref{equivalence1}) is of no practical use.

\subsection{Commuting QTM}
\begin{figure}\label{TQTMfig}
\psfrag{u}{$u$}
\psfrag{-u}{$-u$}
\psfrag{v}{$v$}
\psfrag{N}{$N$}
\psfrag{L}{$L$}
\psfrag{0}{$0$}
\psfrag{T}{$T$}
\psfrag{T'}{$T'$}
\psfrag{QTM}{${\rm QTM}$}
\psfrag{90}{$90^{\circ}$}
\begin{center}
     { \includegraphics[width=8cm]{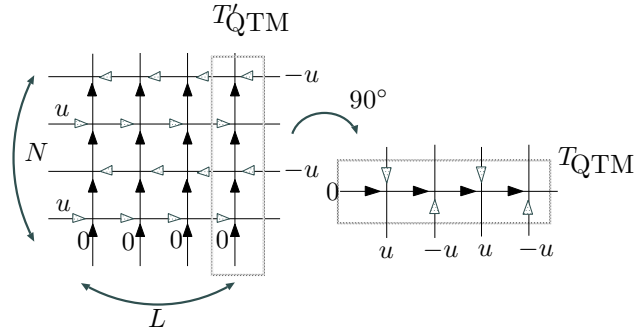}}
	 \caption{The graphical definition of  $T_{\rm QTM}$.}
 \end{center}
 \end{figure}

A crucial observation was made in reference\cite{Suz85}. We start from
the same two-dimensional classical model in fig.~\ref{twoDz}. We consider,
however, the transfer matrix propagating in horizontal direction, that is,
$T'_{\rm QTM}(u) $. 
Equivalently, one can rotate the system by 90${}^\circ $. Then we define
a transfer matrix propagating in vertical direction, $T_{\rm QTM}(u)$
(see fig.~\ref{TQTMfig}). The latter is more convenient for our formulation.

The partition function is then given by,
\begin{equation}
Z_{\rm 1D}(\beta) 
=\lim_{N \rightarrow \infty} 
 {\rm tr}_{V_{\rm Trotter}}  \bigl(T_{QTM} (u) \bigr)^L  \Big |_{ u=-\frac{\beta}{N}}.
\label{equivalence2}
\end{equation}

Let the eigenvalue spectrum of  $T_{\rm QTM}(u)$ be
$\Lambda_0(u) >\Lambda_1(u) \ge \Lambda_2(u) \ge \cdots $. 
Then we have an expansion similar to (\ref{RTRexpansion}) 
\begin{equation}
 {\rm tr}_{V_{\rm Trotter}}  \bigl(T_{QTM} (u) \bigr)^L
 =
 \bigl(\Lambda_0(u) \bigr)^{L} \Bigl( 1+  \bigl( \frac{\Lambda_1 (u) }{\Lambda_0 (u) }\bigr )^{L}  + 
\bigl( \frac{\Lambda_2 (u) }{\Lambda_0 (u) }\bigr )^{L} 
  +\cdots \Bigr ).
\end{equation}

Our physical interest is in the free energy per site $f$ in the
thermodynamic limit $L \rightarrow \infty$.
\begin{align}
f &=  - \frac{1}{\beta } \lim_{L \rightarrow \infty}   \lim_{N \rightarrow \infty}
 \Bigl\{  \ln  \Lambda_0(u)   \nonumber  \\
&\phantom{ccc} +  \frac{1}{L}  \ln \Bigl( 1+  \Bigl( \frac{\Lambda_1 (u) }{\Lambda_0 (u) }\Bigr )^{L}  + 
\Bigl( \frac{\Lambda_2(u) }{\Lambda_0(u) }\Bigr )^{L} 
 +\cdots \Bigr )  \Bigr \} \Big |_{u =-\frac{\beta}{N}}.
\label{freeenergypersite}
\end{align}

\begin{proposition}
The two limits in  (\ref{freeenergypersite}) are exchangeable.
\end{proposition}
We supplement an argument 
which claims that the second term in the second line in 
(\ref{freeenergypersite}) is negligible for $L \rightarrow \infty$.
The previous argument, using the inversion relation (\ref{inversion})
can not be applied directly as the spectral parameter $u$ is already
fixed as $-\frac{\beta}{N}$ in the present problem.

We introduce a slight generalization, a commuting QTM $T_{QTM}(x,u)$, 
by assigning the parameter $ix$ in ``horizontal" direction\cite{Klu92}.
The substitution  $x=0$ recovers the previous results. The precise
definition is shown in (\ref{defQTM}). Hereafter we drop the $u$
dependence as it is always  $-\frac{\beta}{N}$.
Let ${\cal T}_{\rm QTM}(x)$ be the corresponding  monodromy matrix.
Then it is easy to see that monodromy matrices are intertwined by the
same $R$ matrix as in the RTR case,
\begin{equation}
R(x, x') {\cal T}_{\rm QTM}(x) \otimes  {\cal T}_{\rm QTM}(x')
=  {\cal T}_{\rm QTM}(x') \otimes  {\cal T}_{\rm QTM}(x) R(x, x').
\label{RTTQTM}
\end{equation}
This immediately proves the commutativity of ${ T}_{\rm QTM}(x)$ with
different $x$'s.

The most important consequence of introducing $x$ is that we have
the inversion relation in this new ``coordinate",
\begin{equation}
{\Lambda}^{\sharp}_j(x-i)  {\Lambda}^{\sharp}_j(x+i)  =\psi(x, u)
\label{inversionx}
\end{equation}
where we set again
$\Lambda_j(x)=\prod_a \tgh \frac{\pi }{4} (x-\theta_a)
\tgh\frac{\pi }{4}(x+\theta_a)  {\Lambda}^{\sharp}_j (x)$. Note that
${\Lambda}^{\sharp}_j(x)$ also depends on the ``old" spectral parameter
$u$, which is set to be $-\frac{\beta}{N}$ on both sides. The known
function $\psi$ is again independent of $j$. The analysis of the Bethe
ansatz equation associated to the QTM implies that $\theta_ a\sim
\frac{2}{\pi} \ln \frac{4\beta}{\Delta_j}$ for large $\beta$. Then,
proceeding as before, we obtain,
$$
\left  |\frac{\Lambda_j(x)}{\Lambda_0(x) } \right | \sim {\rm e}^{ -\frac{\Delta_j}{\beta} \ch \frac{\pi}{2} x}
 \qquad {\text{thus }} \qquad
\left  |\frac{\Lambda_j(x)}{\Lambda_0(x) } \right |^L  \sim {\rm e}^{ -\frac{\Delta_j L}{\beta} \ch \frac{\pi}{2} x}.
$$
%
 %
%
%
The diagonalization for fixed $N$ clearly shows the gap between the
eigenvalues, which is consistent with the above argument.
Thus, at any finite temperature, the second term in
(\ref{freeenergypersite}) is negligible for $L \rightarrow \infty$.
We then conclude that the formula (\ref{formulaf}) is valid.
 
Although we made use of the integrability of the model in the above
argument, the validity of the formula is actually independent of it.
See the proof in reference\cite{Suz85}.
 
\section{Diagonalization and NLIE}
\subsection{Bethe roots}
Thanks to (\ref{RTTQTM}), one can apply the machinery of the quantum
inverse scattering method,  devised  originally for the diagonalization
of $T_{\rm RTR}$, to the diagonalization of $T_{\rm QTM}$. We skip
the derivation and present only results relevant for our subsequent
discussion\footnote{A technical remark: the vacuum is conveniently chosen
$(+,-,+,-,\cdots)$.}.  
We fix $N$ for a while. Then the eigenvalue of $T_{\rm QTM}$ is given by 
\begin{align}
\Lambda^{(N)}(x) &=    a(x)  \frac{Q(x-2i)} {Q(x)}   +
 d(x) \frac{Q(x+2i)} {Q(x)}    \label{eigenvalue} \\
a(x)&:=  \phi_+(x+2i) \phi_-(x)& 
d(x)&:=  \phi_-(x-2i) \phi_+(x)   \nonumber \\
 Q(x) &:=\prod_{j=1}^m (x-x_j)&
\phi_{\pm} (x)& :=\bigl(\frac{x\pm iu}{ \pm 2i}\bigr )^{\frac{N}{2}}.   \nonumber
\end{align}
The different sets of Bethe roots $\{x_j\}$ correspond to the  different
eigenvalues. They satisfy the Bethe ansatz equation (BAE),
\begin{equation}
 \frac{  a(x_j)  }{d(x_j) } =
 -\frac{Q(x_j +2i)}{Q(x_j-2i)}  \qquad  1\le j \le m.
 \label{BAE}
\end{equation}
For the largest eigenvalue the number of roots $m$ equals $\frac{N}{2}$.
 
To evaluate $f$ via (\ref{formulaf}) we need the largest $\Lambda^{(N)}$
for $N \rightarrow \infty$. This means we must deal with infinitely many
roots in the limit,
which resembles the situation encountered in the evaluation of the
free energy in the thermodynamic limit of a classical 2D model by means of
the RTR transfer matrix.
Still, we would like to comment on the qualitative difference in the
root distribution between such ``standard" case and the problem under
discussion.

\begin{figure}[hbtp]
\centering
{  \includegraphics[width=4cm]{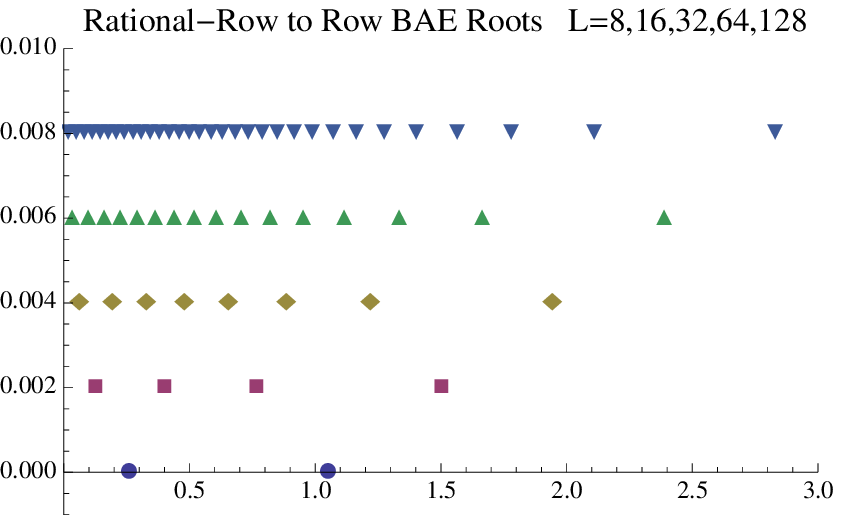} \hspace{1cm}
  \includegraphics[width=4cm]{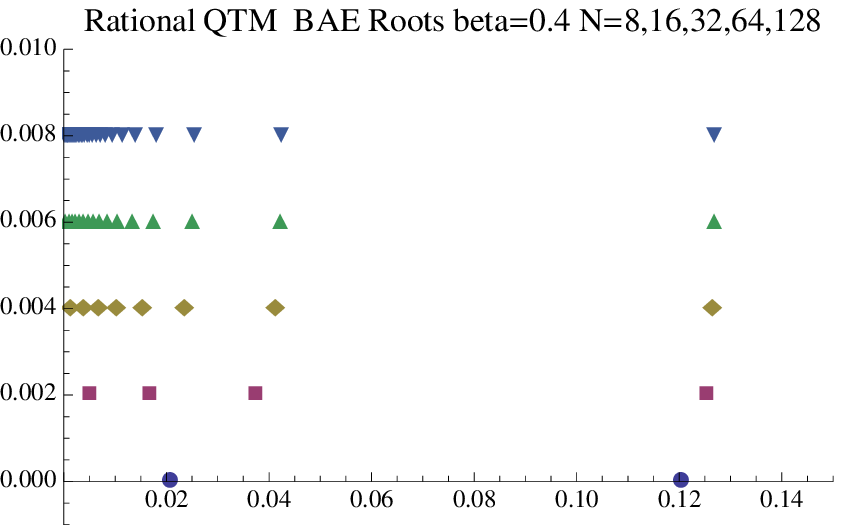} }
\caption{The positive half of BAE roots for RTR (left) and for QTM (right) with system size,  from $8$ (bottom) to $128$ (top).
}
\label{baerootsfig}
\end{figure}

Fig.~\ref{baerootsfig} shows the distribution of the positive half of
BAE roots for the largest eigenvalue of $T_{\rm RTR}$ (left) and
$T_{\rm QTM}$ (right) for various system sizes.
The distribution of RTR roots behaves smoothly for large system size.
The limiting shape of the distribution (the root density function) is
a smooth function satisfying a linear integral equation.
For $T_{\rm QTM}$, on the other hand, a few large roots remain isolated
at almost the same positions as $N$ increases, while close to the origin
more and more Bethe roots cluster. 
%
%

Let us describe this in detail. Using the NLIE technique, we can derive
an approximate BAE equation (see the discussion after (\ref{finiteNLIE})), 
\begin{equation}
\frac{N}{2} \ln \Bigl( - 
\frac{\tgh \frac{\pi}{4} ( x_j -  \frac{\beta}{N} i)    }
           {\tgh \frac{\pi}{4} (x_j +  \frac{\beta}{N} i)   }  \Bigr)\sim  (2 I_j+1) \pi i.
\label{approximateBAE}
\end{equation}
The hole $\theta_a$ corresponds to the branch cut integer $ I_{\rm max} =
\frac{N}{4}-\frac{1}{2}$, and this implies $\theta_a \sim \frac{1}{\pi}
\ln \beta$  for $\beta \gg 1$, which was used in the last subsection.

Near the origin, we set $x_j=\frac{\beta}{N} \widehat{x}_j$, and obtain
the approximate distribution of $\widehat{x}_j$ as an algebraic function, 
$$
\rho(\widehat{x})  = \lim_{N \rightarrow \infty }  \frac{I_{j+1}-I_{j}}{N(\widehat{x}_{j+1}- \widehat{x}_{j})}
 \sim \frac{1}{2 \pi (\widehat{x}^2+1)}.
$$
This differs from the usual root density function which decays exponentially
as $|x| \rightarrow \infty$.
In the original variable, if we take the Trotter limit naively,  
$$
\rho(x)  \sim   \lim_{|u| \rightarrow 0}  \frac{|u|}{2 \pi (x^2+u^2)} 
\sim \frac{1}{2} \delta(x).
$$
Namely the distribution of the BAE roots for $T_{\rm QTM}$ is singular in
the Trotter limit. We thus conclude that the usual root density method
may not be applicable, and we have to devise a different tool. \pn
Let us stress again that the cancellation of of
the order of $N$ many terms in $\ln \Lambda^{(N)}$ is a unique property of
the QTM.
In the RTR case $O(L)$ many terms can survive, and we obtain intensive
quantities (e.g., the free energy per site) only after dividing by $L$.
On the other hand, $\ln \Lambda^{(N)}$ is already an intensive
quantity, as remarked after Theorem \ref{Suzukitheorem}. 
The cancellation is thus vital. The  $O(e^N)$ terms, $a(x), d(x)$ must be canceled
by the denominator $Q(x)$, resulting in $O(1)$ quantities.
According to this point of view, (\ref{eigenvalue}) is (again) not
practical, as the ratios of $O(e^N)$ terms are still present.
Thus, we understand (\ref{eigenvalue}) still as a starting point, not
as the goal.

One must intrinsically deal with a finite size system with coupling
constant $u$ depending on the Trotter size, and then take the Trotter limit.
Such  an attempt was executed first numerically\cite{Koma87} by
extrapolation in $N$. The analytic low temperature expansion was 
performed\cite{SuAkWa90} based on the Wiener-Hopf method.
Below we shall present the most sophisticated approach which utilizes
the commuting QTM in a most efficient manner\cite{Klu93}.

\subsection{Non-linear Integral Equation (NLIE)}
The introduction of the new spectral parameter $x$ plays a fundamental
role. Instead of dealing with the BAE roots directly, we make use of the
analyticity of specially chosen auxiliary functions in the complex $x$
plane.

There are various approaches. One of them is to introduce the fusion
hierarchy of the QTM, which contains the original $T_{\rm QTM}$ as $T_1$.
In place of the BAE one uses the functional relations among fusion
transfer matrices,
\begin{equation}
T_m(x-i)  T_m(x+i)  = \psi_m(x) + T_{m-1}(x) T_{m+1}(x).
\end{equation}
Here $\psi_1$ is nothing but $\psi$ in the inversion relation
(\ref{inversionx}), where the small term $T_0 T_2/\psi_1$ was neglected.
As $\{T_m\}$ constitutes a commuting family, the same relation holds among
the eigenvalues. 
We thus use the same symbol  $T_m$ for the eigenvalue.
After a change of variables, $y_m(x)=  T_{m-1}(x) T_{m+1}(x)/\psi_m(x)$,
one can transform the algebraic equations into integral equations under
certain assumptions on the analyticity of $y_m$. The resultant equations
coincide with the Thermodynamic Bethe Ansatz equations\cite{Klu92,KSS}.
The string hypothesis is thus replaced by an assumption on the analyticity
of $y_m$.   
The coupled set of equations may fix the values of $T_1(x)$. Then
$\Lambda_0=T_1(0)$ yields the free energy per site $f$.   
%
%
A technical problem in this approach is that we must deal with an
infinite number of $y_m$ functions, which requires a truncation of
the equations in an approximate manner. 

Below we shall discuss another approach originally devised in the
context of the evaluation of finite size corrections\cite{KBP}. 
We define the auxiliary function $\mathfrak{a}_N(x)$ by the ratio of the
two terms in $\Lambda^{(N)}(x)$  (\ref{eigenvalue}),
$$
\mathfrak{a}_N(x) =    \frac{d(x)}{a(x)} \frac{Q(x+2i)}{Q(x-2i)}
\qquad
\mathfrak{A}_N(x) = 1+\mathfrak{a}_N(x).
$$
The suffix $N$ is introduced to recall that we are fixing $N$ finite here.
The BAE (\ref{BAE}) is equivalent to the condition 
\begin{equation}
\mathfrak{a}_N(x_j) =-1  \qquad{\rm or}  \qquad    \ln \mathfrak{a}_N(x_j) =(2 I_j+1) \pi i.
\label{quantization}
\end{equation} 
We also note that $\lim_{|x| \rightarrow \infty} \mathfrak{a}_N(x) = 1$
by construction.  

We then adopt the following assumptions for the analytic properties of 
$\mathfrak{A}_N(x)$ corresponding to the largest eigenvalue. They
are supported by numerical calculations.
\begin{enumerate}
\item There are $\frac{N}{2}$ simple zeros of $\mathfrak{A}_N(x)$ on the
real axis. They coincide with the BAE roots. There are additional zeros,
sufficiently far away from the real axis, so that ${\cal C}$ does not
include them inside.
\item The only pole of $\mathfrak{A}_N(x)$ in $\Im x \in [-1,1]$ is
located at $x=iu$ and is of order $\frac{N}{2}$.
\end{enumerate}

Once these assumptions are granted, one immediately derives the following
NLIE,
\begin{equation}
\ln \mathfrak{a}_N(x)=
\ln \frac{\phi_-(x+2i)  \phi_+(x)}{ \phi_+(x+2i)  \phi_-(x)}
 -\int_{{\cal C}} \frac{2}{(x-y)^2+4} \ln \mathfrak{A}_N(y) \frac{dy}{\pi}.
  \label{NLIEbeforelimit}
\end{equation}
The largest eigenvalue $\Lambda$ can be similarly represented by 
\begin{equation}
\ln\! \Lambda^{(N)} (x)\! =\! \ln\bigl(\!\phi_+(x+2i) \phi_-(x-2i) \!\bigr)\!  
 \!+\!\int_{\cal C}
 \frac{ \ln \mathfrak{A}_N(y)}{(x-y)(x-y-2i)} \frac{dy}{\pi}.
  \label{freebeforelimit}
\end{equation}
Note that only the driving term in (\ref{NLIEbeforelimit}) depends on $N$.
We can thus  take the Trotter limit easily,  with $\mathfrak{a}
:=\lim_{N \rightarrow \infty} \mathfrak{a}_N$, and obtain the NLIE in
(\ref{NLIE}) (for $h=0$).
To evaluate the free energy one has to first set $x=0$, then take the
Trotter limit. Or otherwise one meets a spurious divergence. 
%
Then we obtain the expression for the free energy in Theorem \ref{theoremf}.
 
One still needs to make an effort to achieve a high numerical accuracy,
especially at very low temperatures.
The introduction of another pair of auxiliary functions solves this problem.
We define $\bar{\mathfrak{a}}_N, \bar{\mathfrak{A}}_N$ by
$\bar{\mathfrak{a}}_N(x)=(\mathfrak{a}_N(x) )^{-1} $, 
$\bar{\mathfrak{A}}_N(x)=1+ \bar{\mathfrak{a}}_N(x)$.

Numerically one finds that $|\mathfrak{a}_N| \lessgtr 1$ for
$\Im x \gtrless 0 $.
Thus, we use  $\mathfrak{a}_N, \mathfrak{A}_N$ in the upper half plane and
$\bar{\mathfrak{a}}_N, \bar{\mathfrak{A}}_N$ in the lower half plane.
It is straightforward to rewrite (\ref{NLIEbeforelimit}) in the coupled form,
\begin{align}
\!\!\ln \! \mathfrak{a}_N(x)\!\! &=\!\! D^{(N)}_+(x) 
\!\!+\!\!\int_{C_+} \!\! F\!(\!x\!-\!y\!)   \ln \! \mathfrak{A}_N(y) \frac{dy}{2\pi}
\!\! -\!\!\int_{C_-}\!\! F\!(\!x\!-\!y\!)   \ln  \!\bar{\mathfrak{A}}_N(y)  \frac{dy}{2\pi}     \label{finiteNLIE}\\
\!\! \ln  \!\bar{\mathfrak{a}}_N(x) \!\!  &=\!\!   D^{(N)}_-(x)  
\!\!+\!\!\int_{C_-}  F\!(\!x\!-\!y\!)  \ln \!\bar{\mathfrak{A}}_N(y)  \frac{dy}{2\pi} 
\!\!-\!\!\int_{C_+}F(\!x\!-\!y\!)  \ln \! \mathfrak{A}_N(y)  \frac{dy}{2\pi}   \nonumber \\
 \!\!D^{(N)}_{\pm} \!& =\!  
 \frac{N}{2}\! \ln \bigl(  \frac{{\rm th} \frac{\pi}{4}(x+iu)}{ {\rm th} \frac{\pi}{4}(x-iu) }   \bigr),
 \quad
  F(x)\!= \!\int_{-\infty}^{\infty} \!\frac{ {\rm e}^{-ikx}}{1+{\rm e}^{2|k|}} dk,
 \nonumber
 \end{align}
where $C_+ (C_-)$ is a straight contour slightly above (below) the real
axis. In the first (second) equation we understand that $ x \in C_+  (C_-)$.
Note that the convolution terms bring  only minor contributions as they
are defined on those contours where the auxiliary functions are small.   
Therefore the main contributions come from the known functions.
This enables us to perform  numerics with high accuracy.
We can  drop the convolution terms  for the lowest order approximation. 
Thanks to eq.\ (\ref{quantization}) this leads to
eq.\ (\ref{approximateBAE}).
 
Similarly, for the largest eigenvalue we have
\begin{align*}
\!\ln \Lambda^{(N)}(x)\!&=\! \varepsilon^{(N)} (x)
\!\!+\!\!
  \int_{C_+}\!\!  K_+(\!x-x'\!) \! \ln  \!\mathfrak{A}_N (x') \!\frac{dx'}{2\pi}
 \!\!+\!\!
 \int_{C_-} \!\! K_-(\!x-x'\!) \!\ln \!\bar{\mathfrak{A}}_N (x')\! \frac{dx'}{2\pi},  \\
 K_{\pm}(x)&=K(x\pm i), \qquad  K(x)= \frac{\pi}{2\ch \pi x/2}, \\
 \varepsilon^{(N)} & =
 \ln \phi_+(x+2i) \phi_-(x-2i)-\frac{N}{2} \int {\rm e}^{-|k|-ikx} \frac{\sh u k}{k \ch k} dk.
 \end{align*}
 
We obtain the NLIE and the eigenvalue in the Trotter limit by replacing
$\mathfrak{a}_N \rightarrow \mathfrak{a}$ etc.\ and 
$$
 D^{(N)}_{\pm} \rightarrow  
 - \frac{\pi i \beta}{2\sh \frac{ \pi}{2} x}  \quad
\varepsilon^{(N)}   \rightarrow  -\frac{\beta}{2}(1 - \int \frac{1}{1+{\rm e}^{2|k|}} dk).
$$
For the actual calculation, it is even better to deal with
$
\mathfrak{b}(x) := \mathfrak{a}(x+ i)$ and $ \bar{\mathfrak{b}}(x) := \bar{\mathfrak{a}}(x- i)$
so that the singularities of $\ln(1+\mathfrak{b}), \ln(1+\bar{\mathfrak{b}})$
are away from the integration contours. We omit, however, the details.
  
\begin{figure}[hbtp]
\psfrag{chi}{$\chi$}
\psfrag{T}{$T$}
\psfrag{inf}{\tiny{$\infty$}}
\psfrag{XiT}{$\xi T$}
\centering
{\includegraphics[width=5.cm]{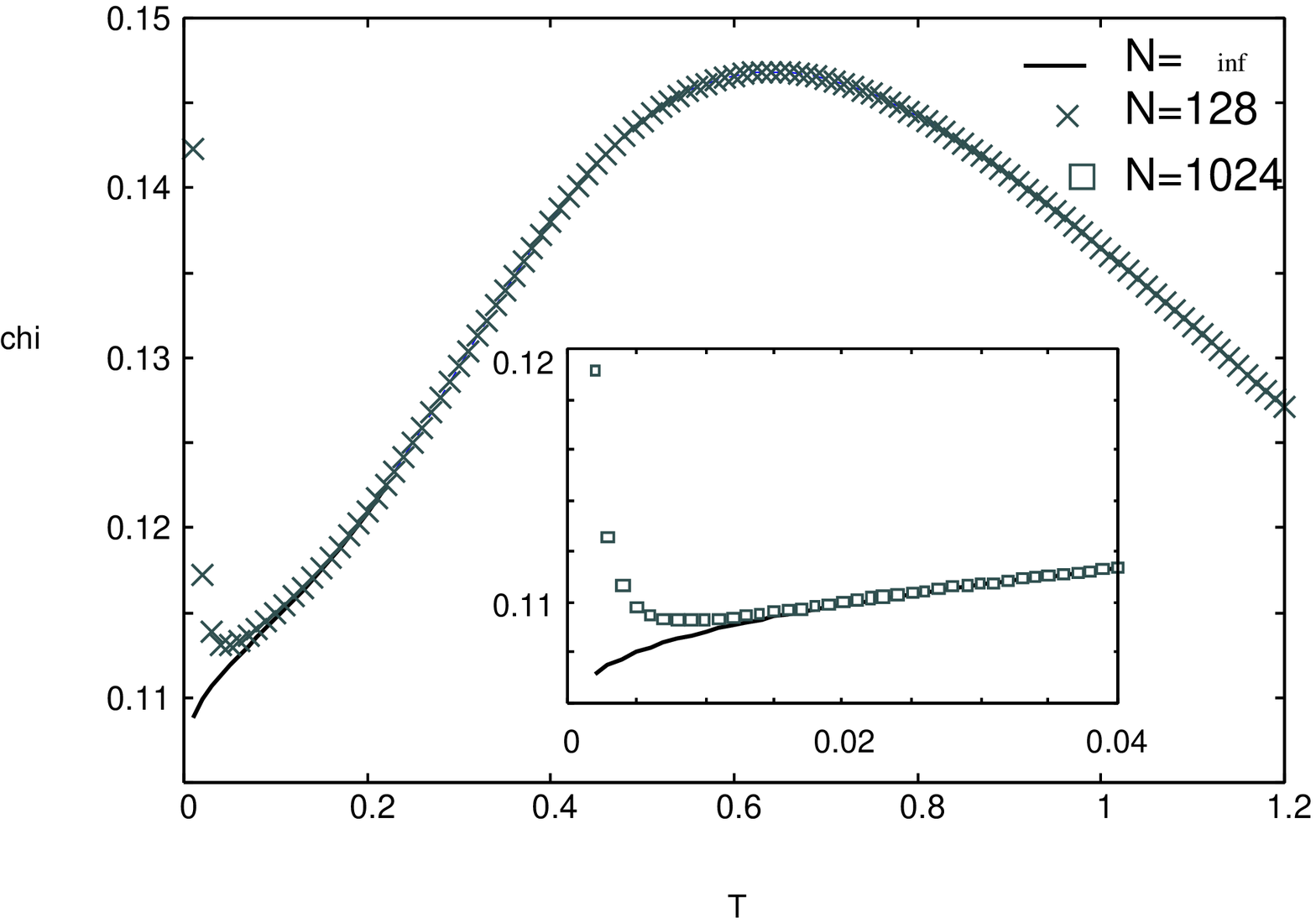}   \hspace{1cm}
\includegraphics[width=5.cm]{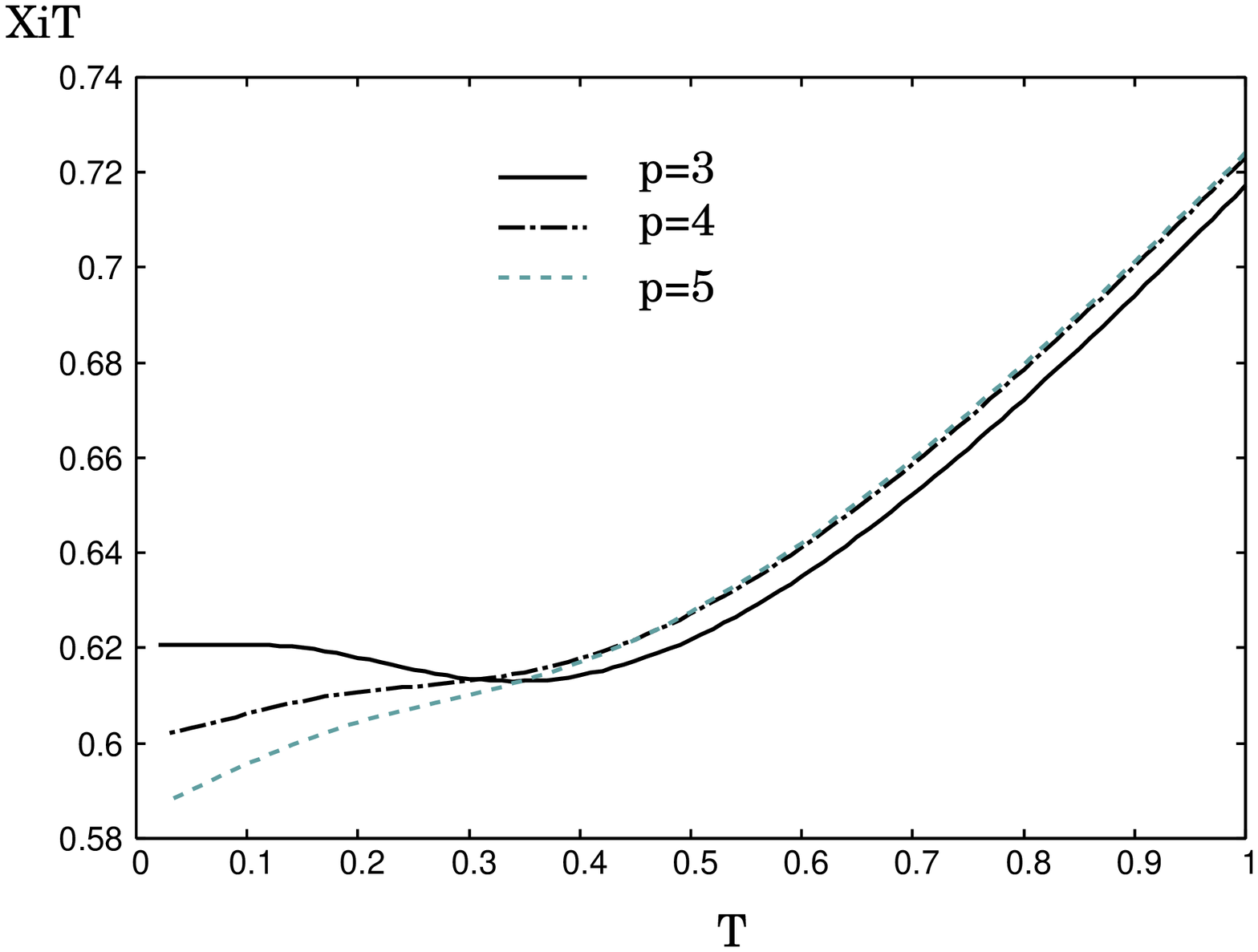}
}
\caption{Left: the susceptibility of the $s=\frac{1}{2}$ XXX model, in the
Trotter limit (solid line) and for fixed Trotter number (crosses:$N=128$,
squares: $N=1024$). Right: a plot of $\xi T$ against temperature for the
XXZ model.
}
\label{bulkplots}
\end{figure}

As a concrete example for the evaluation of bulk quantities we plot
the susceptibility,  $\chi=\partial^2_h f$, in fig.~\ref{bulkplots} (left).
Note that at low temperatures the physical result in the Trotter limit
(solid line) deviates from its finite Trotter number approximation.
The above approach has been successfully applied to many models of
physical relevance\cite{JK,JKS, JSuzE8, FK, JShigherS,  BK, DK}.

The correlation length $\xi$ characterizes the decay of correlation
functions at large distance, e.g.,
\begin{equation}
\langle \sigma^+_x\sigma^-_y \rangle 
 \sim  {\rm e}^{-\frac{|x-y|}{\xi}} \qquad |x-y| \gg 1.
\label{correlationlength}
\end{equation}
It is evaluated from the ratio of the largest and the second largest
eigenvalues of the QTM\cite{SuzukiInoue,SuAkWa90,SSSU,KMSS}. 
For the second largest eigenvalue state, our assumption (1) on
$\mathfrak{A}$ is no longer valid: a pair of holes $\theta_a$ lies on
the real axis, and they are zeros of $\mathfrak{A}$ other than Bethe roots.
Nevertheless, a small modification leads to a set of
equations that fixes the second largest eigenvalue.
The resultant NLIE has a form similar to (\ref{finiteNLIE}) containing,
however, additional inhomogeneous terms. Fig.~\ref{bulkplots} (right)
shows a plot of $\xi T$ against temperature for the XXZ model
with $q={\rm e}^{i\frac{\pi}{p}}$ for $p=3,4,5$\cite{SSSU}.  

When $h \ne 0$, we have to replace $a(x), d(x)$ in (\ref{eigenvalue}) by
${\rm e}^{-\beta h/2}a(x)$, ${\rm e}^{\beta h/2}d(x)$. Then we add
$\beta h$ $( -\frac{\beta h}{2})$ to the rhs of (\ref{NLIEbeforelimit})
((\ref{freebeforelimit})). Also $D^{(N)}_{\pm}$  must be replaced by
$D^{(N)}_{\pm}\pm \frac{\beta h}{2}$.

Before closing this section, we would like to mention another formulation
of thermodynamics also based on the QTM\cite{ShiroTak}. It is described
by a NLIE for $\Lambda_j$ directly and allows one to efficiently
calculate high temperature expansions. The good numerical accuracy in the low
temperature region is, however, hard to achieve. 
Moreover, we point out that the equation is the same for any eigenvalue.
Thus, one should know a priori good initial values in order to select the 
convergence to the desired eigenvalue.

\section{DME (density matrix elements) at finite temperatures}
The deep understanding of a model requires ample knowledge of its correlation
functions. 
We would therefore like to go beyond their asymptotic characterization by
the correlation length $\xi$ (\ref{correlationlength}).

The evaluation of correlation functions has been defying many
challenges in the past. Considerable progress was made only recently
for the $T=0$ correlations, based on vertex operators\cite{JMMN}, on
the $q$KZ equation\cite{JimboMiwa96} and on QISM\cite{KMT}.
The third approach is the most relevant for our purpose.  
For $T=0$ it first requires the solution of the ``inverse problem", that
is, one has to represent the spin operators in terms of the QISM
operators $A(u), B(u),C(u), D(u)$. Then, by algebraic manipulations,
one obtains the correlation functions as combinatorial sums of expectation
values of QISM operators, which are finally converted into (multiple)
integrals.

At first glance, the case $T>0$ seems far more difficult, as one expects
that a summation of the contributions from all excited states is necessary.
We argue here that, as above, the QTM helps us to avoid this summation
and that, moreover, one does not have to solve the ``inverse problem" 
within the QTM framework.  
The combinatorics, on the other hand, can be done in parallel to $T=0$,
because the QISM algebra is the same in both cases.  
%
 
\begin{figure}
\psfrag{a1}{$\alpha_1$}
\psfrag{a2}{$\alpha_2$}
\psfrag{a3}{$\alpha_3$}
\psfrag{b1}{$\beta_1$}
\psfrag{b2}{$\beta_2$}
\psfrag{b3}{$\beta_3$}

\psfrag{(u,x)}{$(u,x)$}
\psfrag{x}{$ix$}
\begin{center}
     { \includegraphics[width=3.5cm]{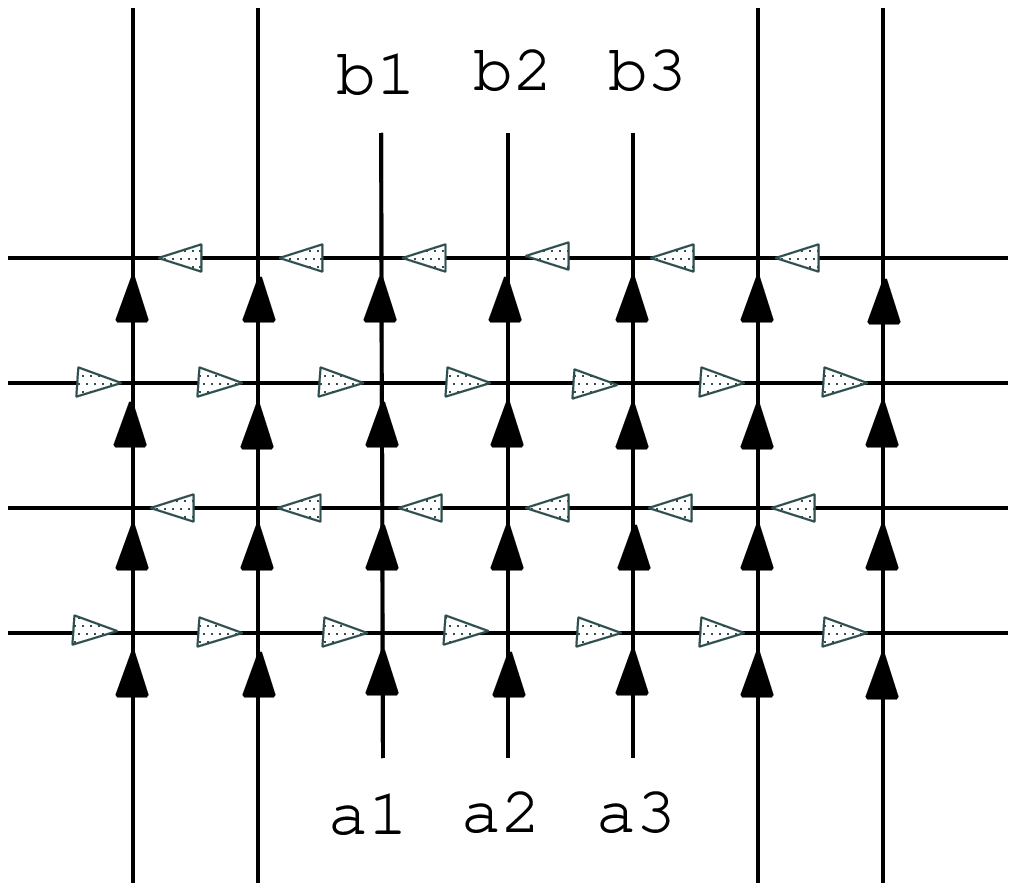}
	\hspace{1cm}
	 \includegraphics[width=3.5cm]{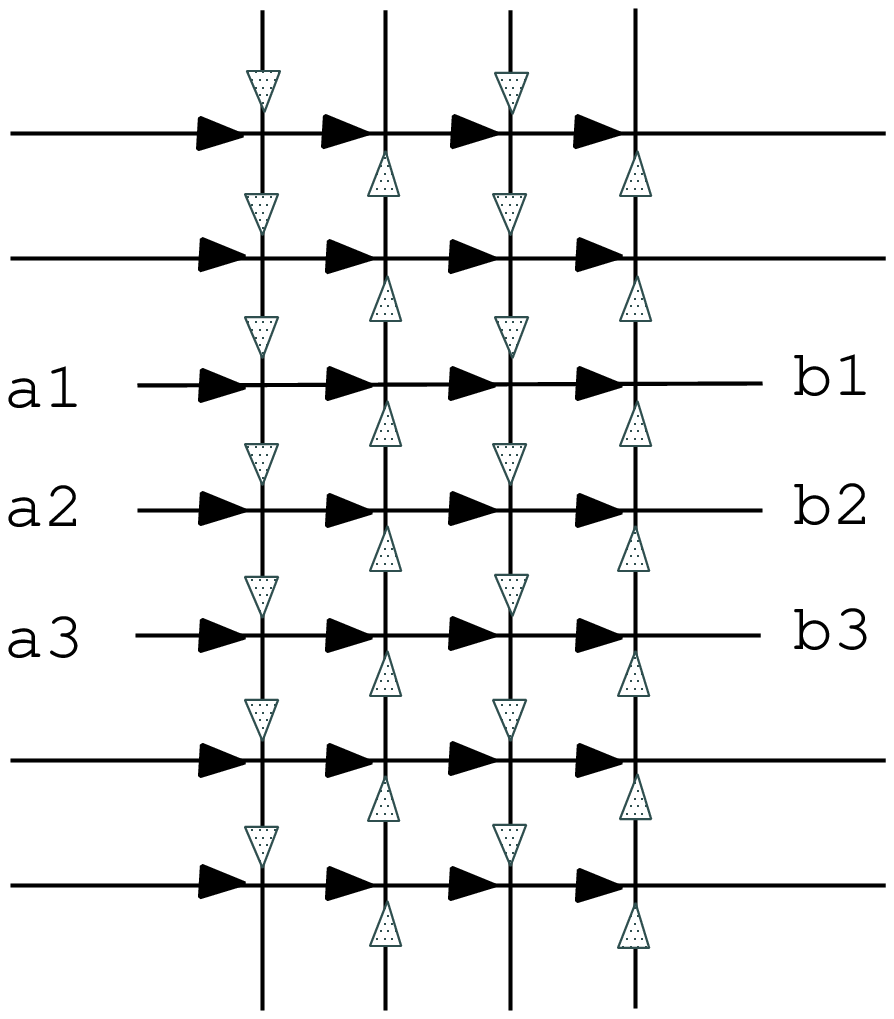}
	 }
	 \caption{Left: A graphically representation of $D^{\alpha_1,\alpha_2,\alpha_3}_{\beta_1,\beta_2,\beta_3}$.
	 Right: The same figure rotated by 90${}^\circ$.}
 \label{DMEfig}
 \end{center}
 \end{figure}

Let us explain why we can bypass the inverse problem in the QTM formalism.
This can be most quickly done in a graphical manner.
To be specific, we need to evaluate DME,
$$
D^{\alpha_1\cdots \alpha_m }_{\beta_1 \cdots \beta_m}
:=  \langle E^{\alpha_1}_{\beta_1} \cdots E^{\alpha_m}_{\beta_m}  \rangle
=\frac{{\rm tr}_{V_{\rm phys}} {\rm e}^{-\beta {\cal H}}  E^{\alpha_1}_{\beta_1} \cdots E^{\alpha_m}_{\beta_m} }
{{\rm tr}_{V_{\rm phys}}  {\rm e}^{-\beta {\cal H}}}.
$$
Using the logic of section 2, we can represent ${\rm e}^{-\beta {\cal H}}$
by a ``2D partition function". Therefore
$D^{\alpha_1\cdots \alpha_n }_{\beta_1 \cdots \beta_n}$ can be
represented by a modified 2D partition function:
Start from the $N\times L$ classical system (fig. \ref{twoDz}) with
periodic boundaries in both directions. Then cut $n$ successive
vertical bonds at the bottom row, and fix the variables at both sides
of the cut. As we are adopting PBCs in the vertical
direction this is equivalent to fixing the configuration of
$n$ successive bonds at the top and at the bottom. See fig.~\ref{DMEfig}
(left).

As previously, we rotate the lattice by 90${}^\circ$. See fig.~\ref{DMEfig} (right).
Obviously we can write $D^{\alpha_1\cdots \alpha_n }_{\beta_1 \cdots \beta_n}$
in terms of elements of the monodromy matrix ${\cal T}_{\rm QTM}(x)$.
By introducing
independent spectral parameters $\xi_i$
we obtain
\begin{equation*}
\Bigl( D \Bigr)^{\alpha_1,\cdots, \alpha_n}_{\beta_1,\cdots, \beta_n}
 (\xi_1,\cdots,\xi_n)
 = 
 \frac{\langle \Phi_0 | ({\cal T}_{\rm QTM})^{\alpha_1}_{\beta_1} (\xi_1) 
   \cdots ({\cal  T}_{\rm QTM})^{\alpha_n}_{\beta_n} (\xi_n) | \Phi_0 \rangle}
   {\langle \Phi_0 | T_{\rm QTM}(\xi_1) \cdots T_{\rm QTM} (\xi_n) | \Phi_0 \rangle}.
\end{equation*}
Here $\Phi_0$ denotes the largest eigenvalue state of the QTM, which is
given by acting with $B$ operators on the vacuum.
 %
As  any $({\cal T}_{\rm QTM})^{\alpha_i}_{\beta_i} (\xi_i)$
is  represented by a QISM operator, we reach  an 
expression for DME purely in terms of QISM
operators without solving the inverse problem.

At the same time,
the problem for $T>0$ is not so simple in view of the analyticity. We
consider $D^{++}_{++}$ as a concrete example. After employing the standard
QISM algebra, one obtains 
\begin{eqnarray*}
&D^{++}_{++}(\xi_1,\xi_2) \times \mathfrak{A}(\xi_1) \mathfrak{A}(\xi_2)=
\Big(  
\sum_{j,k} 
 \frac{ (x_{k}-\xi_2) (x_{j}-\xi_1-2i) }{\xi_{2,1} (x_j-x_k-2 i)} 
\begin{vmatrix}
w_{j,1}& w_{k,1} \\
w_{j,2}& w_{k,2}
\end{vmatrix}   \\
&- 
\frac{\xi_{1,2}+2i}{\xi_{1,2}}  \sum_{j}   \frac{ (x_{j}-\xi_2)  }{(x_j-\xi_2+2 i)} w_{j,1}  
-\frac{\xi_{2,1}+2i}{\xi_{2,1}}
  \sum_j  \frac{ (x_{j}-\xi_1)  }{(x_j-\xi_1+2 i)} w_{j,2}
+1
 \Bigr).
\end{eqnarray*}
Here $x_j$ denotes a BAE root and $\mathfrak{A}$ is the auxiliary function.
We introduced  $w_{j,k}$ in order to deal with the ratio of inner products
of wave functions. $w_{j,k}$ is characterized by a simple algebraic
relation.
Note that the above algebraic expression is formally identical for $T=0$
and $T>0$: one only has to replace $x_j$ and $ w_{j,k}$ for $T=0$ by those
for $T>0$.
  
In the case $T=0$ there are several simplifications. First, the auxiliary
function is by construction trivial, $\mathfrak{A}=1$. Second, we can
introduce the root density function in the thermodynamic limit. Then the
algebraic relation for $w_{j,k} \rightarrow g(x_j,\xi_k)$    is 
solved with the explicit result $g(x,\xi)=\frac{1}
{4 \ch \frac{\pi}{2}(x-\xi+i) }$.
\begin{eqnarray*}
&D^{++}_{++}(\xi_1,\xi_2)=
\Big(  
\int dx dx' 
 \frac{ (x'-\xi_2) (x-\xi_1-2i) }{\xi_{2,1} (x-x'-2 i)} 
\begin{vmatrix}
g(x,\xi_1) &g(x',\xi_1)  \\
g(x,\xi_2) & g(x',\xi_2) 
\end{vmatrix}   \\
&- 
\frac{\xi_{1,2}+2i}{\xi_{1,2}}  \int dx   \frac{ (x-\xi_2)  }{(x-\xi_2+2 i)} g(x,\xi_1) 
-\frac{\xi_{2,1}+2i}{\xi_{2,1}}
  \int dx  \frac{ (x-\xi_1)  }{(x-\xi_1+2 i)}  g(x,\xi_2) 
+1
 \Bigr)  .
\end{eqnarray*}
Third, we can freely move the integration contours. Every time it passes
the singularity of $g(x)$, it brings extra contributions and they cancel
the ``tails" (the 2nd to the 4th terms above). We finally obtain
\begin{equation}
\!\!D^{++}_{++}(\xi_1,\xi_2)\!\!=\!\!
\int_{-\infty}^{\infty}\!\!dx dx' 
 \frac{ (x'\!-\!\xi_2\!+\!i) ( x\!-\!\xi_1\!- \!i) }{\xi_{2,1} (x-x'-2 i)} 
\begin{vmatrix}
g(x\!+\!i,\xi_1) &g(x'\!+\!i,\xi_1)  \\
g(x\!+\!i,\xi_2) & g(x'\!+\!i,\xi_2) 
\end{vmatrix}.
\label{TzeroD}
\end{equation}
Without such a compact expression, it is hard to proceed further.

On the other hand, $\mathfrak{A}$ is quite non-trivial for $T>0$.
As noted previously, we can not resort to the root density function.
The explicit form of $w_{j,k}$ in the Trotter limit is thus unknown.
The most significant difference is that the integration contour is already
fixed for $T>0$. Thus, we cannot apply the above trick to swallow tails
into the ground state.

Nevertheless, with an appropriate choice of a further auxiliary function
$G(x,\xi)$, it was shown that a compact multiple integral representation,
similar to (\ref{TzeroD}) is also possible for $T>0$\cite{GKS,GHS},
$$
\!\!D^{++}_{++}(\xi_1,\xi_2)\!\!=\!\!
\int_{\cal C} \frac{dx}{{\mathfrak A}(x)} \! \int_{\cal C} \frac{dx'}{{\mathfrak A}(x')}
 \frac{ (x-\xi_1-2i) ( x'-\xi_2) }{4 \pi^2 \xi_{1,2} (x-x'-2 i)} 
\begin{vmatrix}
G(x,\xi_1) &G(x',\xi_1)  \\
G(x,\xi_2) & G(x',\xi_2) 
\end{vmatrix}.
$$
The formula for any other DME is similarly known.

It is a big progress to obtain the multiple integral representation
for DMEs. 
The representation is, however, not yet
optimal. Although one can use it for the numerical analysis at sufficiently
high temperatures, it suffers from numerical inaccuracy at low
temperatures\cite{BG,TsuboiHT}. We thus would like to reduce it to (a sum
of) products of single integrals.

The factorization of DME at $T=0$ has been performed
by brute force,  with the extensive use of the shift of contour technique
\cite{BoosKorepin, TakahashiG}.
Based on studies of the $q$KZ equation, a hidden Grassmannian structure 
behind  DME has been found\cite{BJMST}.
It naturally explains the factorization of the multiple integral
formula through the nilpotency of operators. 
The explicit form of DME consists of two pieces, the algebraic part,
evaluated from a matrix element of $q$-oscillators, and the transcendental
part, related to the spinon-$S$ matrix.

\begin{figure}
\begin{center}
    { \includegraphics[width=10cm]{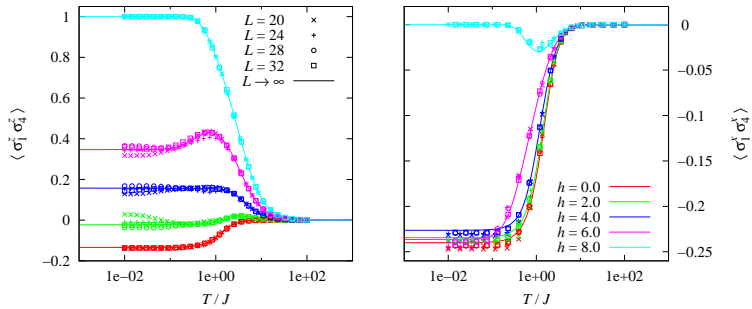}}
	 \caption{The plots of  $\langle \sigma^z_1  \sigma^z_4 \rangle$  (left) and 
	  $\langle \sigma^x_1  \sigma^x_4 \rangle$ (right)
	  for $\Delta=\frac{1}{\sqrt{2}}$ with various magnetic fields
	  by NLIEs (continuous line).}
 \label{XXZzzcorr}
\end{center}
\end{figure}

On the other hand, we still do not have a finite temperature analogue of
the $q$KZ equation. We are nevertheless able to factorize the multiple
integrals for small segments\cite{WuppertalG}. 
The explicit results also consist of two parts. Surprisingly the algebraic part remains identical
to the $T=0$ case, while the transcendental part can be interpreted as a
proper finite temperature analogue to the spinon-$S$ matrix. 
This finally enables us to perform an accurate numerical analysis of the
correlation functions\cite{WuppertalG2}. We show plots of $\langle
\sigma^z_1  \sigma^z_4 \rangle$ for $\Delta=\frac{1}{\sqrt{2}}$ with
various magnetic fields in fig.~\ref{XXZzzcorr}. The results from a brute
force calculation are also plotted, which supports the validity of our
formula.
Such high accuracy calculation can clarify the quantitative nature of
interesting phenomena such as the quantum-classical
crossover\cite{FabriciusMcCoy}.  \pn
Recently a proof of the existence of factorization of the DMEs for $T>0$
was obtained, again by using the Grassmannian structure \cite{HGSIII}.
See also the further development\cite{BG09} in this direction based on
NLIEs.

\section{Summary and discussion}
We presented a brief review on the recent progress with the exact
thermodynamics of 1D quantum systems. The QTM is found to be an efficient
tool, and it offers a framework to evaluate quantities of physical
interest, including DME. The NLIE combines into
the framework nicely, yielding high accuracy numerical results.

The factorization of the multiple integral formula at $T>0$ is yet to
be further explored. It seems e.g.\ quite plausible that the $q$KZ
equation could be extended to finite temperatures. This might be an
important next step.

There are certainly many interesting questions left open. For example,
can we have the QTM formulation starting from a continuum system?
What is the generalization of the multiple integral formula to models
with higher spin? The study of such questions is underway.

\section*{Acknowledgments}
The authors take pleasure in dedicating this review to Professor Tetsuji
Miwa on the occasion of his sixtieth birthday. They thank the organizers
of ``Infinite Analysis 09" for their  warm  hospitality.


\begin{thebibliography}{9}
 
\bibitem{Gaudin71}
M.~Gaudin, Phys. Rev. Lett. {\bf 26} (1971) 1301.

\bibitem{TakSuz}
M.~Takahashi and M. Suzuki, Prog. Theor. Phys. {\bf 48} (1972) 2187-2209.

\bibitem{Bax8v}
R. J. Baxter, Ann. Phys. {\bf 76} (1973) 1, ibid, 25, ibid 48.

\bibitem{Suz85}
M.~Suzuki, Phys. Rev. B {\bf 31} (1985) 2957-2965.

\bibitem{SuNaWa92}
For earlier reviews on the QTM and finite size correction method, see
e.g.\ J.~Suzuki, T.~Nagao and M.~Wadati, Int. J. Mod. Phys. B {\bf 6} (1992)
1119-1180.
A. Kl{\"u}mper, Lecture Notes in Phys. {\bf 645} (2004) 349-379,
or the review chapters in the book by F. H. L. Essler,
H. Frahm, F. G{\"o}hmann, A. Kl{\"u}mper and V. E. Korepin, 
``The One-Dimensional Hubbard Model" (Cambridge Press 2005).

\bibitem{Jimbo}
M. Jimbo, Lett. Math. Phys. {\bf 10} (1985) 63-69.

\bibitem{Klu93}
A. Kl\"umper, Z. Phys. B {\bf 91} (1993) 507-519.

\bibitem{Koma87}
T.~Koma, Prog. Theor. Phys. {\bf 78} (1987) 1213-1218.

\bibitem{SuAkWa90}
J.~Suzuki, Y.~Akutsu and M.~Wadati, J. Phys. Soc. Jpn. {\bf 59} (1990)
2667-2680.

\bibitem{Klu92}
A.~Kl\"umper, Ann. Physik  (Lpz.) {\bf 1} (1992) 540.


\bibitem{KSS}
A. Kuniba, K. Sakai and J. Suzuki, Nucl. Phys. {\bf B 525} [FS] (1998)
597-626.

\bibitem{KBP}
A. Kl\"{u}mper, M. T. Batchelor and P. A. Pearce, J. Phys. A {\bf 24}
(1991) 3111-3133.
				
\bibitem{JK}
G.~J{\"u}ttner and A.~Kl{\"u}mper, Euro. Phys. Lett. {\bf 37} (1997) 335-340.

\bibitem{JKS}
G.~J{\"u}ttner, A.~Kl{\"u}mper and J.~Suzuki, Nucl. Phys. B {\bf 486}
(1997) 650-574, J. Phys. A {\bf 30} (1997) 1881-1886, Nucl. Phys. B
{\bf 512} (1998) 581-600, Nucl. Phys. B {\bf 522} (1998) 471-502.
  
\bibitem{JSuzE8}
J. Suzuki, Nucl. Phys. B {\bf 528} (1998) 683-700.

\bibitem{FK}
A. Fujii and A. Kl{\"u}mper, Nucl. Phys. B {\bf 546} (1999) 751-764.
  
\bibitem{JShigherS}
J. Suzuki, J. Phys. A {\bf 32} (1999) 2341-2359.
    
\bibitem{BK}
M. Bortz and A.  Kl{\"u}mper, Eur. Phys. J. B {\bf 40} (2004) 25-42.

\bibitem{DK}
J. Damerau and A. Kl{\"u}mper, JSTAT (2006) P12014.

\bibitem{SuzukiInoue}
M. Suzuki and M. Inoue, Prog. Theor. Phys. {\bf 78} (1987) 787.

\bibitem{SSSU}
K. Sakai,  M. Shiroishi, J. Suzuki and Y. Umeno, Phys. Rev. B {\bf 60}
(1999) 5186-5201.

\bibitem{KMSS}
A. Kl{\"um}per, J. R.  Reyes Martinez, C. Scheeren, M. Shiroishi,
J. Stat. Phys. {\bf 102} (2001)  937-951.

\bibitem{ShiroTak}
M. Shirosihi and M. Takahashi, Phys. Rev. Lett. {\bf 89} (2002) 117201(4pp).
Z. Tsuboi, J. Phys. A {\bf 37} (2004) 1747-1758.

\bibitem{JMMN}
M. Jimbo, K. Miki, T. Miwa, A. Nakayashiki, Phys. Lett. A {\bf 16}
(1992) 256-263.

\bibitem{JimboMiwa96}
M. Jimbo and T. Miwa, J. Phys. A {\bf 29} (1996) 2923-2958.

\bibitem{KMT}
N. Kitanine, J. M. Maillet and V. Terras, Nucl. Phys. B {\bf 567}
(2000) 554-582.  N. Kitanine, J. M. Maillet, N. A. Slavnov,
V. Terras, Nucl. Phys. B {\bf 641} (2002) 487-518.

\bibitem{GKS}
F. G{\"o}hmann, A. Kl{\"um}per and A. Seel,
J. Phys. A {\bf 37} (2004) 7625-7651.

\bibitem{GHS}
F. G{\"o}hmann, N. Hasenclever and A. Seel, JSTAT (2005) P10015.

\bibitem{BG}
M. Bortz and F. G{\"o}hmann,
Eur. Phys. J. B {\bf 46} (2005) 399-408.

\bibitem{TsuboiHT}
Z. Tsuboi,
Physica A {\bf 377} (2007) 95-101.

\bibitem{BoosKorepin}
H. E. Boos and V. E. Korepin, J. Phys. A {\bf 34} (2001) 5311-5316.
H. E. Boos,  V. E. Korepin and F. A. Smirnov, Nucl. Phys. B {\bf 658} (2003)
417-439, J. Phys. A {\bf 37} (2004) 323-336.
  
\bibitem{TakahashiG}
K. Sakai, M. Shiroishi, Y. Nishiyama and  M. Takahashi,
Phys. Rev. E {\bf 67} (2003) 06510(R) (4pp).
G. Kato, M. Shiroishi, M. Takahashi, K. Sakai, J. Phys. A {\bf 36} (2003)
L337-L344.  J. Sato, M. Shiroishi, M. Takahashi JSTAT (2006) P017.

\bibitem{BJMST}
H. Boos, M. Jimbo, T. Miwa, F. Smirnov, Y. Takeyama,
Comm. Math. Phys. {\bf 261} (2006) 245-276, Annales Henri-Poincare
{\bf 7} (2006) 1395-1428, Comm. Math. Phys {\bf 272} (2007) 263-281,
Comm. Math. Phys. {\bf 286} (2009) 875-932.

\bibitem{WuppertalG}
H. E. Boos, F. G{\"o}hmann, A. Kl{\"u}mper and J. Suzuki,
JSTAT 0604 (2006) P001, J. Phys. A {\bf 40} (2007) 10699-10728.

\bibitem{WuppertalG2}
H. E. Boos, J. Damerau, F. G{\"o}hmann, A. Kl{\"u}mper and J. Suzuki
and A. Wei{\ss}e, JSTAT 0808 (2008) P08010.

\bibitem{FabriciusMcCoy}
K. Fabricius and B. M. McCoy, Phys. Rev. B {\bf 59} (1999) 381-386.
K. Fabricius, A.  Kl{\"u}mper and B. M. McCoy, Phys. Rev. Lett.
{\bf 82} (1999) 5365-5368.

\bibitem{HGSIII}
 M. Jimbo, T. Miwa and  F. Smirnov,
 J. Phys.  A {\bf 42} (2009) 304018 (31pp).
 
\bibitem{BG09}
H. Boos and F.  G{\"o}hmann, J. Phys.  A{\bf 42} (2009) 315001 (27pp).

\end{thebibliography}
\end{document}